\documentstyle[aps,preprint,psfig]{revtex}
\begin{document}
\begin{center}
{\bf THE GLAUBER MODEL AND THE HEAVY ION REACTION CROSS SECTION}
\end{center}
\begin{center}
Prashant Shukla \\
{\it Nuclear Physics Division \\
Bhabha Atomic Research Center, Mumbai 400 085, India} \\
\end{center}

\begin{abstract}

  We reexamine the Glauber model and calculate the total reaction 
cross section as a function of energy in the low and intermediate energy 
range, where many of the corrections in the model, are effective.
  The most significant effect in this energy range is by the 
modification of the trajectory due to the Coulomb field. The modification 
in the trajectory due to nuclear field is also taken into account in 
a self consistent way. 
  The energy ranges in which particular corrections
are effective, are quantified and it is found that when the center of 
mass energy of the system becomes 30 times the Coulomb barrier, 
none of the trajectory modification to the Glauber model is really 
required. 
  The reaction cross sections for light and heavy systems,
right from near coulomb barrier to intermediate energies 
have been calculated. The exact nuclear densities and free 
nucleon-nucleon (NN) cross 
sections have been used in the calculations. The center of mass 
correction which is important for light systems, has also been taken 
into account. 
  There is an excellent agreement between the calculations 
with the modified Glauber model and the experimental data. 
This suggests that 
the heavy ion reactions in this energy range can be explained by 
the Glauber model in terms of free NN cross sections without 
incorporating any medium modification.

\end{abstract}

\newpage
\section{Introduction}

 The Glauber model (GM) \cite{GLAUBER} has been employed for
describing heavy ion reactions at high energies.
It is a semiclassical model picturing the nuclei moving 
in a straight path along the collision direction, and gives the
nucleus-nucleus interaction \cite{KAROL} in terms of interaction
between the constituent nucleons (NN cross section)
and nuclear density distributions. It is a well established model
for high energies and has been applied to heavy ion collision for
describing a number of reaction processes
(See e.g. \cite{BLAIZ,WONG,SHUKNUC}). 
One of the most important physical quantities characterizing the
nuclear reactions is the total reaction cross section.
 It is very useful for extracting information about the nuclear sizes. 
The Glauber model has been successively used to get the radii of radioactive
nuclei from measured total reaction cross sections \cite{TANIHATA}.

   At low energies, the straight line trajectory is modified due
to the Coulomb field between two nuclei.
The Coulomb modified Glauber model (CMGM) \cite{CMGM,CHAGUP} consists
of replacing
the eikonal trajectory at an impact parameter $b$, with the eikonal
trajectory at the corresponding distance $r_c$ of closest approach
in the presence of the Coulomb field.
Later the non eikonal nature of the trajectory
around $r_c$, has also been taken into account  \cite{SHUKLA}.
 Replacing $r_c$ by the distance $r_{cn}$ of closest approach in the
presence of both the Coulomb and the nuclear field gives the 
Coulomb plus Nuclear modified Glauber
model (CNMGM) \cite{SHUKLA,CHA}. This model (CMGM/CNMGM) has been widely
used in recent literature \cite{WARNER,AHMAD,CAI,PNU}. 
 Warner et. al. \cite{WARNER} have shown in 
their calculations for light projectiles on various targets 
that trajectory modifications are minor. 
Let us remark that the energies considered by them is 
much above the Coulomb barrier. 
 The Coulomb modified Glauber model has also been used at very 
high energies \cite{AHMAD} where trajectory modifications may
be ineffective. There have been various prescriptions \cite{CAI,PNU} to 
modify the NN cross sections due to nuclear density.
  Most of the work reported earlier \cite{CMGM,CHAGUP,CHA,SHUKLA,CAI,PNU}
has been done using Gaussian densities. The reaction cross section
is sensitive to surface density of the colliding nuclei. 
Depending on how well one has produced the surface texture
with the Gaussian form, it will make a 5 to 10 \% change in
the reaction cross section over that done with realistic 
densities \cite{CHARNEW}.

   In the light of this, we reexamine the various trajectory corrections in
the Glauber model and calculate the total reaction cross section as a 
function of energy upto 50 times the Coulomb barrier.
  We quantify the energy range in which a particular correction
is effective and find that when the center of mass energy 
of the system becomes 30 times the Coulomb barrier, the above 
modifications to GM are insignificant.
 Thus, the energy at and above which the results of
CNMGM and GM coincide depends on the 
Coulomb barrier between the two nuclei and will be different
for a light and heavy system.
 In the present work we use exact nuclear densities and 
free NN cross sections.
 The center of mass correction which is important for light systems 
has also been taken into account. 
 Comparison of the experimental data with our calculations are 
presented in the plots of $\sigma_R(E)$ vs. $E$. 


\section{The Glauber model}
 Consider the collision of a projectile nucleus $A$ on a target
nucleus $B$.
 The probability for occurrence of a nucleon-nucleon collision
when the nuclei $A$ and $B$ collide at an impact
parameter $\bf b$ relative to each other is given by \cite{WONG,SHUKNUC}
\begin{equation}\label{wong}
T(b) \bar \sigma_{NN} = \int \rho_A^z({\bf b}_A) d{\bf b}_A \,
 \rho_B^z({\bf b}_B) d{\bf b}_B \,
 t({\bf b-b_A+b_B}) \,\, \bar \sigma_{NN}.
\end{equation}

 Here $\rho_A^z({\bf b}_A)$ and $\rho_B^z({\bf b}_B)$
are the z-integrated densities of projectile and target nuclei
respectively. $t({\bf b}) d {\bf b}$ is the probability for having
a nucleon-nucleon collision within the transverse area element
$d {\bf b}$ when one nucleon approaches at an impact parameter
${\bf b}$ relative to another nucleon.
All these distribution functions are normalized to one.
Here $\bar \sigma_{NN}$ is the average total nucleon nucleon cross section.

 There can be upto $A\times B$ collisions. The probability
of occurrence of $n$ collisions will be
\begin{equation}
 P(n,b) ={AB \choose n} (1-s)^n (s)^{AB-n} .
\end{equation}
Here, $s=1-T(b) \bar \sigma_{NN}$.
The total probability for the occurrence of at least one 
NN collision in the collision of $A$ and $B$ at an impact parameter 
${\bf b}$ is
\begin{eqnarray}
{d \sigma_{R} \over db} = \sum_{n=1}^{AB} P(n,b)
 = 1 - s^{AB}.
\end{eqnarray}
  The total reaction cross section $\sigma_R$ can be written as
\begin{equation}
\sigma_{R}=2 \pi \int b db \left( 1-s^{AB} \right).
\end{equation}
From this one can read the scattering matrix as
\begin{eqnarray}
|S(b)|^2 = s^{AB}=(1-T(b)\bar \sigma_{NN})^{AB}.
\end{eqnarray}
In the optical limit, where a nucleon of the projectile undergoes
only one collision in the target nucleus can be written as
\begin{eqnarray}\label{sb}
|S(b)|^2 \simeq \exp(-T(b)\bar \sigma_{NN} AB).
\end{eqnarray}
The scattering matrix can be defined in terms of eikonal
phase shift $\chi(b)$ as
\begin{eqnarray}\label{sb2}
S(b)= \exp\left(-i\chi (b)\right).
\end{eqnarray}
If we assume $\bar \alpha_{NN}$ to be the ratio of real to imaginary 
part of NN scattering amplitude,
the eikonal phase shift can be obtained as

\begin{eqnarray}\label{rf8}
\chi (b) ={1\over 2}\bar \sigma_{NN} (\bar \alpha_{NN}+i) \, AB \,\,T(b).
\end{eqnarray}
Here $\bar \alpha_{NN}$ will not be directly used for calculating the
reaction cross section but will come through the correction in the
trajectory due to the nuclear field.
Once we know the phase shift and thus the scattering matrix, we can
calculate the reaction cross section and also the 
elastic scattering angular distribution. In the present work we
restrict ourselves to the calculations of reaction cross section 
only.

In momentum space $T(b)$ is derived as
\begin{eqnarray}\label{tbmom}
T(b) = {1\over 2\pi}
       \int J_0(qb) S_{A}({\bf q}) S_{B}(-{\bf q})  f_{NN}(q) q dq.
\end{eqnarray}
Here $S_{A}(q)$ and  $S_{B}(-q)$ are
the Fourier transforms of the nuclear densities
and $J_0(qb)=1/2\pi\int \exp(-qb \cos \phi) d\phi$ is the
cylindrical Bessel function of zeroth order.
The function $f_{NN}(q)$ is the Fourier transform of the
profile function $t({\bf b})$ and gives the $q$ dependence of
NN scattering amplitude.
 The profile function $t({\bf b})$ for the NN scattering can be
taken as delta function if the nucleons are point particles.
In general it is taken as a Gaussian function of width $r_0$ as
\begin{equation}
t({\bf b}) = {\exp(-b^2/2r_0^2) \over 2\pi r_0^2}.
\end{equation}
Thus,
\begin{eqnarray}
f_{NN}(q) = \exp(-r_0^2 q^2/2).
\end{eqnarray}
Here, $r_0$ is the range parameter and has a weak dependence on energy
(see for discussions\cite{AHMAD}). We use $r_0=0.6$ fm \cite{CHAGUP}
in the energy range 2 to 200 MeV.

  The average NN scattering parameter $\bar \sigma_{NN}$
is obtained in terms of pp cross section 
$\sigma_{pp}$ and nn cross section $\sigma_{nn}$, the parameterized
forms for which are available in Ref.~\cite{CHAGUP}.
For projectile energies below 10 MeV/nucleon,  
we use the prescription given in 
Ref.~\cite{CHAGUP} for computing $\bar \alpha_{NN}$.
For energies above 10 MeV/nucleon, we take the parameterized forms from 
the work of Ref.~\cite{GREIN}. These forms produce the elastic 
scattering data for various systems satisfactorily.

\subsection{Optical potential in the Glauber model}

  The usual Glauber optical limit phase shift function is given by
\begin{equation}\label{usual}
 \chi(b)= -\frac{1}{\hbar v}\intop_{-\infty}^{\infty}
          V_N({\bf r}) dz .
\end{equation}
 Comparing Eq.~(\ref{usual}) with Eq.~(\ref{rf8}), the optical potential 
$V_N(r)$ in the momentum representation can be identified as,

\begin{eqnarray}\label{vr}
V_N({\bf r})&=&- \hbar v \,\, \bar \sigma_{NN}
   (\bar \alpha_{NN}+i)\,\,{A B \over 2(2\pi)^3}
  \int e^{-i {\bf q}.{\bf r}} S_A({\bf q}) S_B({\bf -q})
   f_{NN} ({\bf q}) d^3 q , \nonumber \\
&=&- \hbar v \,\, \bar \sigma_{NN}(\bar \alpha_{NN}+i)\,\,{A B \over (2\pi)^2}
   \int j_0(qr) S_A({\bf q}) S_B({\bf -q})
    f_{NN} ({\bf q}) q^2 dq .
\end{eqnarray}
Here $j_0(qr)$ is the spherical Bessel function of zeroth order.

\subsection{Trajectory modifications in the Glauber model}
  The basic assumption in the Glauber model is the description
of the relative motion of the two nuclei in terms of straight line
trajectory. For low energy  heavy ion
reactions, the straight line trajectory is assumed at the distance
of closest approach $r_{c}$
calculated under the influence of the
Coulomb potentials for each impact parameter $b$ as given by,
\begin{equation}
 r_c=(\eta +\sqrt{\eta^2+ b^2 k^2})/k,
\end{equation}
which is a solution of the following equation without the
nuclear potential $V_N(r)$
\begin{equation}\label{rcn}
E-\frac{Z_PZ_T e^2}{r}-\frac{\hbar^2 k^2}{2\mu}\frac{b^2}{r^2}
         - {\rm Re} V_N(r) = 0.
\end{equation}
Here $\eta=Z_P Z_T e^2/\hbar v$ is the dimensionless Sommerfeld parameter.
The CMGM \cite{CMGM,CHAGUP} consists of replacing
the eikonal trajectory at an impact parameter $b$ with the eikonal
trajectory at the corresponding distance $r_c$ of closest approach.
The distance of closest approach $r_{cn}$ in the presence of both
the Coulomb and nuclear potential \cite{SHUKLA} is obtained by solving
Eq.~(\ref{rcn}) numerically, where $V_N(r)$ has been derived  
from Eq.~(\ref{vr}). By replacing $b$ by $r_{cn}$, CNMGM is obtained.
  The non eikonal trajectory \cite{SHUKLA}
around $r_c$ in the presence of the Coulomb field is represented by
$r^2 = r_{c}^2+ (C+1) z^2$, where the quantity $C$  is given by
\begin{equation}
C=\frac{\eta}{kb^2} r_{c}.
\end{equation}
Thus in the CMGM, $T(b)$ will be simply replaced by
$T(r_c(b))/\sqrt{(C+1)}$.

\subsection{The nuclear densities}

  The two parameter fermi (2pf) density is given by
\begin{equation}
\rho(r)=\frac{\rho_0}{1+\exp(\frac{r-c}{d})},
\end{equation}
  where $\rho_0=3/\left(4\pi c^3(1+\frac{\pi^2d^2}{c^2})\right)$.
Thus, the momentum density can be derived \cite{GUPTAZ}as
\begin{equation}
S(q)=\frac{8\pi\rho_0}{q^3} \frac{z e^{-z}}{1-e^{-2z}}
     \left( \sin x  \frac{z(1+e^{-2z})}{1-e^{-2z}}-x \cos x \right).
\end{equation}
  Where $z=\pi dq$ and $x=cq$. Here $d$ is the diffuseness and
$c$, the half value radius in terms of rms radius $R$ for
the 2pf distribution is calculated by 
$c=(5/3R^2-7/3 \pi^2 d^2 - 5/3 r_p^2)^{1/2}$.
Here $r_p$ is the proton radius.
 Equation~(\ref{tbmom}) can be solved numerically for this density 
and the overlap integral can be extracted.
 In the present work, the density parameters have been taken
from the compilation of measured charge 
density distributions \cite{JAGER,VRIES} and are tabulated 
in Table~\ref{table1}.

  For lighter nuclei such as $^{12}$C and $^{16}$O, the densities
are given in the from of harmonic oscillator (HO) densities as
\begin{equation}
\rho(r)=\rho_0 \left( 1+ \alpha {r^2\over R^2} \right)
          \exp(-{r^2\over R^2}),
\end{equation}
where $\rho_0= 1/\left[ 1+ 1.5 \alpha \right]$.
The momentum density is given by
\begin{equation}
S(q)=\rho_0 \left( 1+ 1.5 \alpha - 0.25 \alpha q^2 R^2 \right)
          \exp\left(-{q^2 R^2 \over 4}\right).
\end{equation}

{\bf Center of mass correction:}

  For lighter nuclei such as $^{12}$C and $^{16}$O we also take 
into account the corrections due to Center of motion. Such a 
correction for harmonic oscillator wave functions is given in 
Ref.\cite{FRANCO}. With this the corrected density will become
\begin{equation}
S(q)=\rho_0 \left( 1+ 1.5 \alpha - 0.25 \alpha q^2 R^2 \right)
      \exp\left(-{q^2 R^2 \over 4}\right)
       \exp\left({q^2 R^2 \over 4 A}\right).
\end{equation}

\begin{table}[h]
\caption{Density parameters for various nuclei used in the
present work}
\label{table1}
\begin{center}
\begin{tabular}{|c|c|c|c|c|}
\cline{1-5}
$A$ &$Z$ & Density & $d/\alpha$  &  $R$  \\
  &      &  form   &  fm         &  fm   \\
\cline{1-5}
 12 &  6 &HO & 1.082 &  1.692   \\
 16 &  8 &HO & 1.544 &  1.833   \\
 28 & 14 &2pf& 0.537 &  3.150   \\
 40 & 20 &2pf& 0.563 &  3.510   \\
 58 & 28 &2pf& 0.560 &  3.823   \\
 90 & 40 &2pf& 0.550 &  4.274   \\
208 & 82 &2pf& 0.546 &  5.513   \\
\cline{1-5}
\end{tabular}
\end{center}
\end{table}

\section{Results and discussions}
  The reaction cross sections for light and heavy systems 
right from near coulomb barrier to intermediate energies 
have been calculated. We calculate the total reaction cross section 
as a function of energy upto 50 times the Coulomb barrier. 
The data for various systems and their references are given in 
Table~\ref{table2}, \ref{table3} and \ref{table4}. 
 The Coulomb barrier is calculated from
expression $V_C= Z_P Z_T 1.44/ (1.5(A^{1/3} + B^{1/3}))$. 
The reduced mass $M$ is defined as $M=AB/(A+B)$.

\begin{table}[h]
\caption{Reaction cross section data for $^{12}$C on various 
targets along with their references}
\label{table2}
\begin{center}
\begin{tabular}{|c|c|c|c|}
\cline{1-4}
System  & $E_{\rm lab}/A$  &  $\sigma_R$ &  Ref.  \\
        &  MeV/nucleon           &     mb      &        \\
\cline{1-4}
$^{12}$C + $^{12}$C
&  9.33  &    1444.00 & \cite{KOX}      \\$V_C/M=1.258$
& 15.00  &    1331.00 & \cite{SAHM}     \\
& 25.00  &    1296.00 & --''--     \\
& 30.00  &    1315.00 & \cite{CHAUVIN}  \\
& 35.00  &    1259.00 & \cite{SAHM}     \\
\cline{1-4}
$^{12}$C + $^{40}$Ca
&15.0  &  2165.0 &  \cite{SAHM}  \\    $V_C/M=2.186$
&25.0  &  2030.0 &  --''--  \\
&30.0  &  2014.0 &  --''--  \\
\cline{1-4}
$^{12}$C + $^{90}$Zr
&10.0 &  2219.0 & \cite{BALL}    \\   $V_C/M=3.214$
&15.0 &  2297.0 & \cite{SAHM}    \\
&25.0 &  2415.0 & --''--    \\
&35.0 &  2528.0 & --''--    \\
\cline{1-4}
$^{12}$C + $^{208}$Pb
&  8.0  &   1754.0 & \cite{SAHM}  \\  $V_C/M=5.068$
& 15.0  &   2873.0 & --''--  \\
& 25.0  &   3236.0 & --''--  \\
& 35.0  &   3561.0 & --''--  \\
\cline{1-4}
\end{tabular}
\end{center}
\end{table}

\begin{table}[h]
\caption{Reaction cross section data for $^{16}$O on various 
targets along with their references}
\label{table3}
\begin{center}
\begin{tabular}{|c|c|c|c|}
\cline{1-4}
System  & $E_{\rm lab}$/A  &  $\sigma_R$ &  Ref.  \\
        &  MeV/nucleon           &     mb      &        \\
\cline{1-4}
$^{16}$O + $^{16}$O
& 2.0 &    73.86 &  \cite{BASS} \\ $V_C/M=1.524$
& 3.0 &  1136.00 &   --''--  \\
& 4.0 &  1300.00 &  --''-- \\
& 5.0 &  1395.00 &  --''-- \\
&  9.0625 &  1650.00 &\cite{KHOA} \\
& 15.625  &  1664.00 & --''--\\
& 21.875  &  1639.00 & --''--\\
& 30.000  &  1655.00 & --''--\\
\cline{1-4}
$^{16}$O + $^{28}$Si
&  2.0625  &   509.0    & \cite{CRAMER} \\ $V_C/M=1.90$
&  2.3750  &   765.6    & --''--\\
&  3.1250  &  1341.0    & --''--\\
&  3.4375  &  1451.0    & --''--\\
&  4.125   &  1626.0    & --''--\\
&  5.0625  &  1777.0    & --''--\\
&  8.9000  &  2013.0    & --''--\\
& 13.45    &  2067.0    & --''-- \\
& 94.00    &  1757.0    & \cite{CHA}    \\
\cline{1-4}
\end{tabular}
\end{center}
\end{table}

\begin{table}[h]
\caption{Reaction cross section data for $^{16}$O on various 
targets along with their references}
\label{table4}
\begin{center}
\begin{tabular}{|c|c|c|c|}
\cline{1-4}
System  & $E_{\rm lab}/A$  &  $\sigma_R$ &  Ref.  \\
        &  MeV/nucleon     &     mb      &        \\
\cline{1-4}
$^{16}$O + $^{58}$Ni
&  2.50   &     48.2  & \cite{KEELEY}\\ $V_C/M=2.683$
&  2.75   &    288.7  & --''--\\
&  3.00   &    539.4  & --''--\\
&  3.75   &    994.5  & --''--\\
&  4.375  &   1282.0  & --''--\\
&  5.00   &   1485.0  & --''--\\
&  6.25   &   1765.0  & --''--\\
&  7.50   &   1953.0  & --''--\\
\cline{1-4}
$^{16}$O+$^{208}$Pb
&  5.00   &   124.1 & \cite{KIM}  \\ $V_C/M=5.019$
&  5.50   &   566.6 & --''--  \\
&  6.00   &   939.6 & --''--  \\
&  6.50   &  1171.0 & \cite{BAECH}  \\
&  8.093  &  2023.0 & \cite{BALL} \\
& 12.00   &  2847.0 & --''-- \\
& 19.54   &  3452.0 & \cite{SATCH} \\
& 94.00   &  3600.0 & \cite{ROUSSEL}\\
\cline{1-4}
\end{tabular}
\end{center}
\end{table}

\begin{figure}[h]
\centerline{\psfig{figure=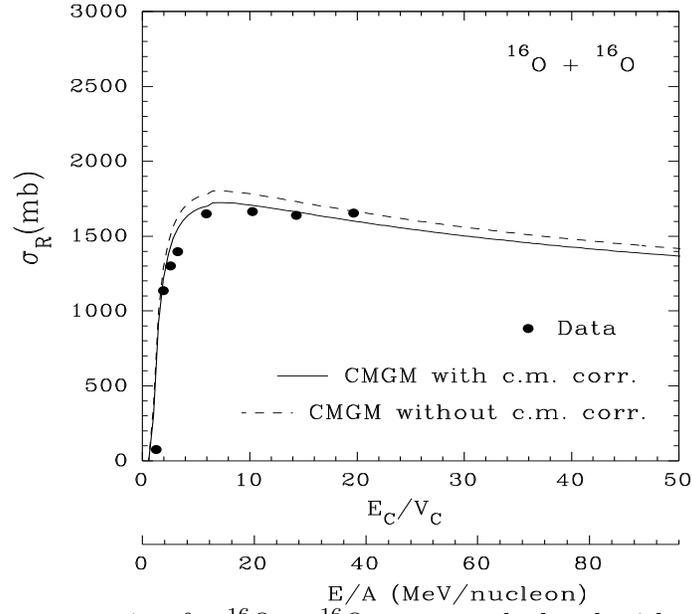,width=9cm,height=8cm}}
\caption{Reaction cross section for $^{16}$O + $^{16}$O system
     calculated with the Coulomb modified
   Glauber model (CMGM) with and without center of mass (c.m.) correction.}
\label{f1}
\end{figure}

\begin{figure}[h]
\centerline{\psfig{figure=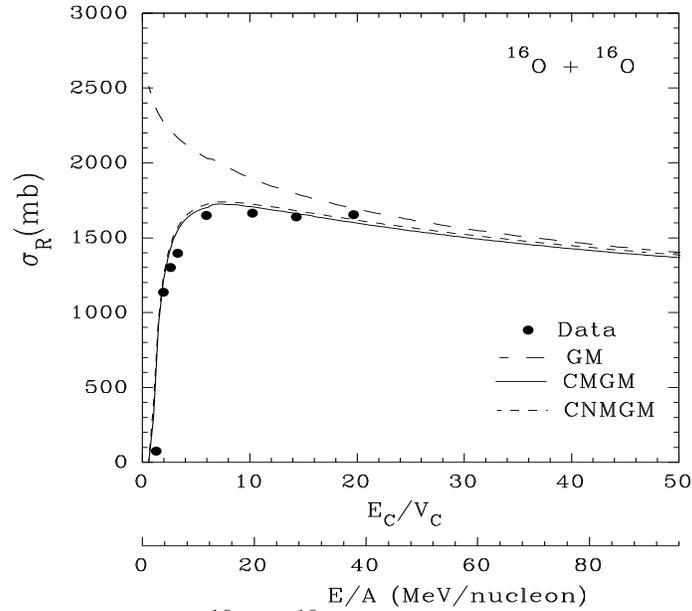,width=9cm,height=8cm}}
\caption{Reaction cross section for $^{16}$O + $^{16}$O system
     calculated with the Glauber model (GM), the Coulomb modified
   Glauber model (CMGM) and the Coulomb plus nuclear modified
   Glauber model (CNMGM) along with the data.}
\label{f2}
\end{figure}

\begin{figure}[h]
\centerline{\psfig{figure=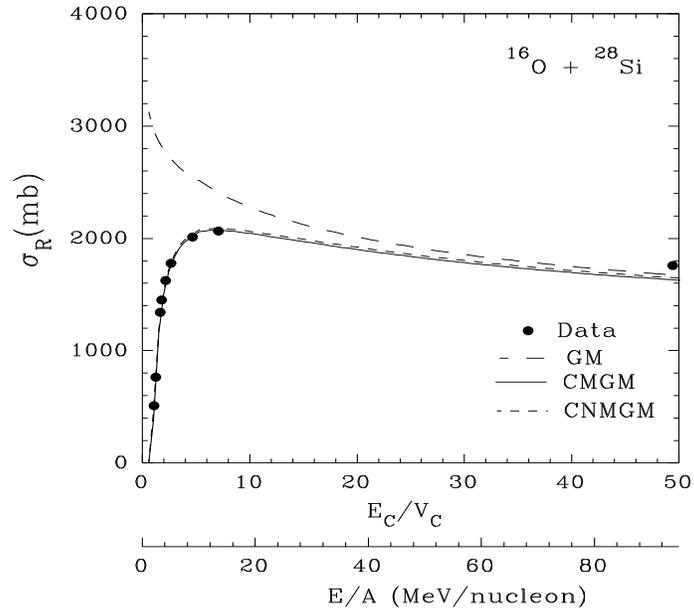,width=9cm,height=8cm}}
\caption{Same as Fig.~2 but for $^{16}$O + $^{28}$Si system.}
\label{f3}
\end{figure}

\begin{figure}[h]
\centerline{\psfig{figure=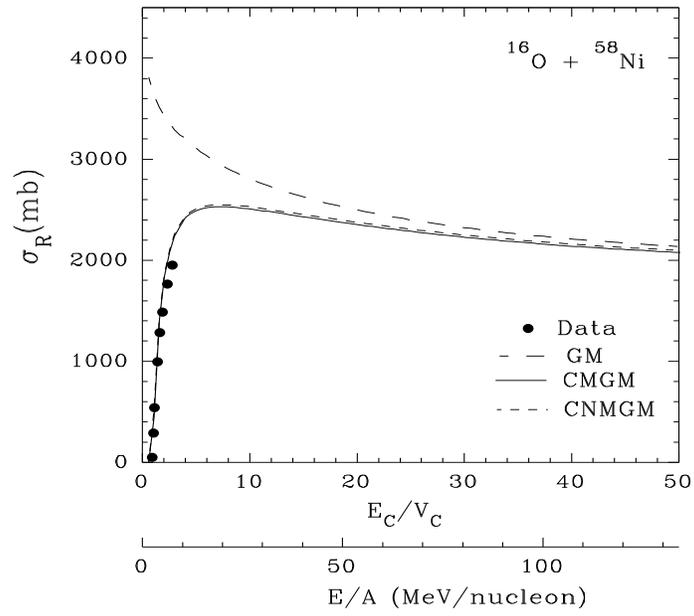,width=9cm,height=8cm}}
\caption{Same as Fig.~2 but for $^{16}$O + $^{58}$Ni system.}
\label{f4}
\end{figure}

\begin{figure}[h]
\centerline{\psfig{figure=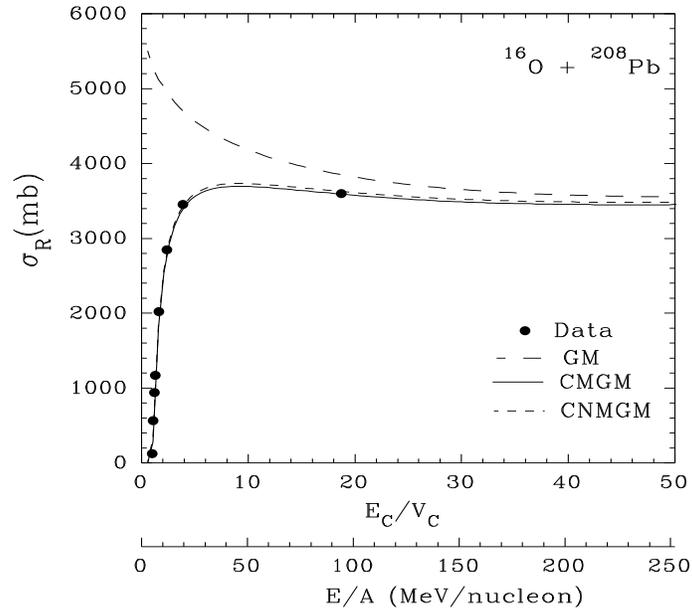,width=9cm,height=8cm}}
\caption{Same as Fig.~2 but for $^{16}$O + $^{208}$Pb system.}
\label{f5}
\end{figure}

\begin{figure}[h]
\centerline{\psfig{figure=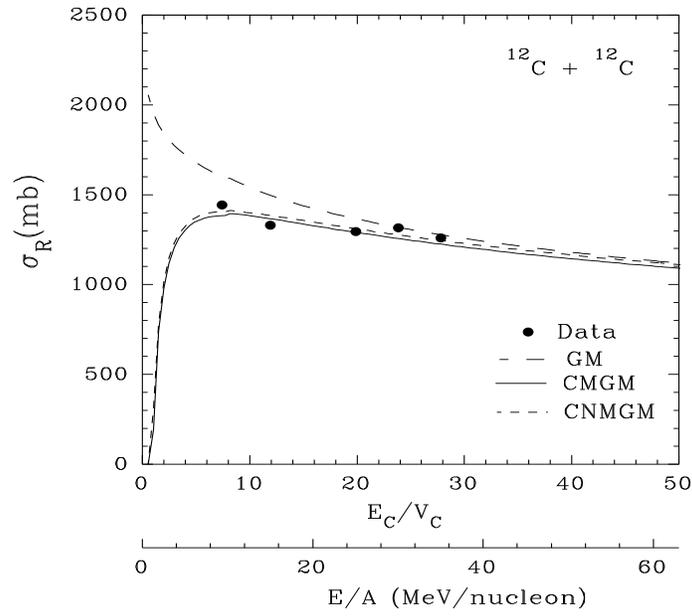,width=9cm,height=8cm}}
\caption{Same as Fig.~2 but for $^{12}$C + $^{12}$C system.}
\label{f6}
\end{figure}

\begin{figure}[h]
\centerline{\psfig{figure=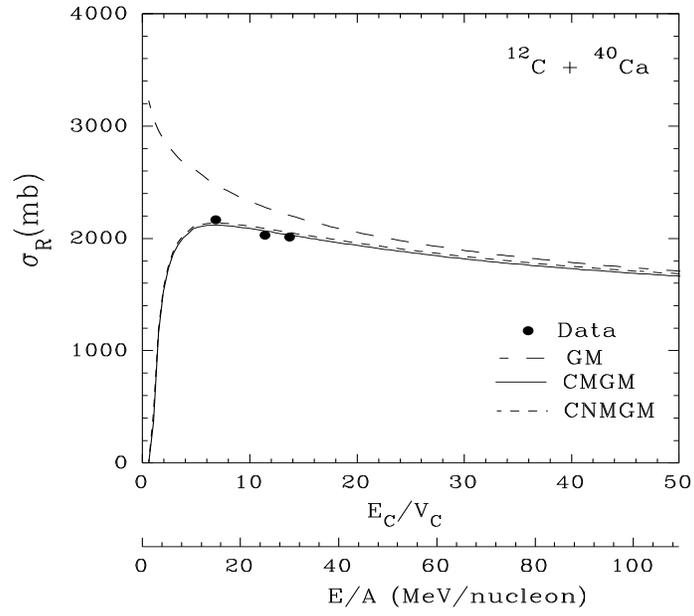,width=9cm,height=8cm}}
\caption{Same as Fig.~2 but for $^{12}$C + $^{40}$Ca system.}
\label{f7}
\end{figure}

\begin{figure}[h]
\centerline{\psfig{figure=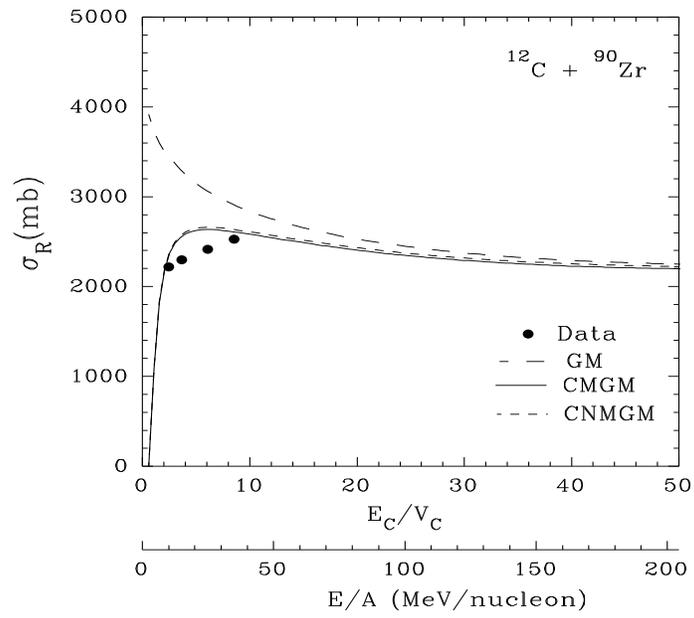,width=9cm,height=8cm}}
\caption{Same as Fig.~2 but for $^{12}$C + $^{90}$Zr system.}
\label{f8}
\end{figure}

\begin{figure}[h]
\centerline{\psfig{figure=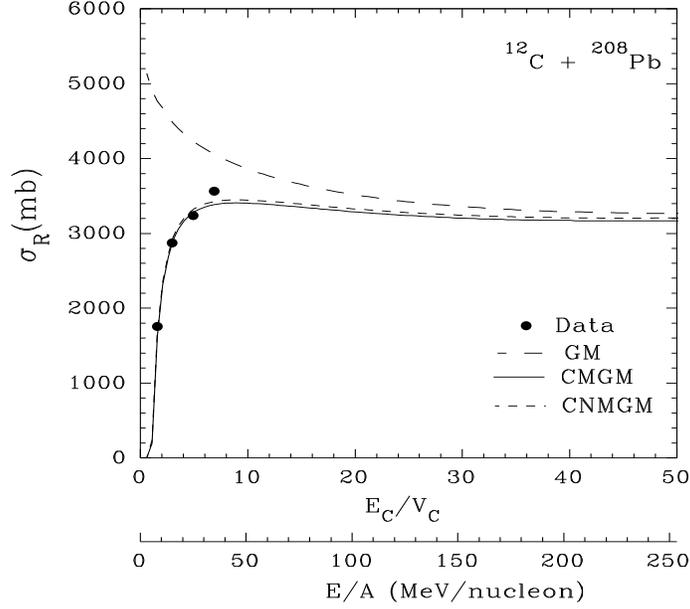,width=9cm,height=8cm}}
\caption{Same as Fig.~2 but for $^{12}$C + $^{208}$Pb system.}
\label{f9}
\end{figure}

Figure~(\ref{f1}) shows the reaction cross section for 
$^{16}$O + $^{16}$O system calculated with CMGM 
with and without center of mass (c.m.) correction. This correction reduces
the reaction cross section for lighter systems.
 Figures~(\ref{f2}) to (\ref{f5}) show the
reaction cross section for $^{16}$O on various  targets 
as a function of center of mass energy divided by the Coulomb
barrier, calculated with the Glauber model (GM), the Coulomb modified
Glauber model (CMGM) and the Coulomb plus nuclear modified
Glauber model (CMNGM) along with the data. 
 Figures~(\ref{f5}) to (\ref{f9}) show the
reaction cross section for $^{12}$C on various targets.
The center of mass correction has been taken into account in all 
the calculations.
  It can been seen from all the figures that the most significant 
effect in this energy range, is by the trajectory modification due 
to the Coulomb field. It is very significant upto energies $6 V_C$.
The modification in the trajectory due to the nuclear field tries to
bring the results closer to GM.
 We universally find that when the center of mass energy 
of the system becomes 30 times the Coulomb barrier, the calculations 
with the CNMGM match with GM within 2 to 3\%.
 Thus, the energy at and above which the results of
CNMGM and GM coincide depends on the 
Coulomb barrier between the two nuclei and not on the 
energy per nucleon of the projectile.
  Further, it is observed that there is an excellent agreement between 
the calculations from the modified Glauber model and the experimental data. 
In contrast to \cite{CAI,PNU} the present study suggests that the heavy ion
reactions in this energy range can be explained by 
the Glauber model in terms of free NN cross sections without 
incorporating medium modification.
 
  There are higher order corrections \cite{WALLACE,VARMA} to the 
optical limit phase shift function $\chi(b)$. 
These corrections are important at higher
$b$ but tend to become smaller at larger $b$ \cite{VARMA} and 
are less significant for total reaction cross section which depend
mostly upon peripheral collisions. They may become important 
for differential cross sections away from forward direction (which 
probe collisions at smaller $b$), increasingly at higher
energies. The Gaussian densities \cite{VARMA} which produce only 
surface textures of the nuclear density are not adequate 
and realistic densities are to be used (as done in \cite{AHMAD})
to calculate these corrections.

\section{Conclusions}
 To summarize, we have reexamined the various trajectory corrections in
the Glauber model and calculated the total reaction cross section as a 
function of energy upto 50 times the Coulomb barrier.
  The most significant effect in this energy range is by the 
trajectory modification due to the Coulomb field. The modification 
in the trajectory due to nuclear field is also taken into account.
We derive $V_N(r)$ consistently from Eq.~(\ref{vr}).
  We quantify the energy range in which a particular correction
is effective and find that when the center of mass energy 
of the system is about 30 times the Coulomb barrier, no trajectory 
modification to GM is really required. 
 Exact nuclear densities and free NN cross sections have been used
in the calculation. The center of mass correction which is important 
for light systems has also been taken into account. 
  There is an excellent agreement between our calculations 
including all the modifications discussed in the manuscript,
and the experimental data. In contrast to \cite{CAI,PNU}
the present study suggests that the heavy ion
reactions in this energy range can be explained by 
the Glauber model in terms of free NN cross sections without 
invoking any medium modification. It implies that the density effects 
on the NN cross sections are either absent or very minor.
One possible explaination for this 
is the fact that for heavy ions, the contribution
to the reaction cross section comes from the surface region
where the density is very small.

\acknowledgements
 The author acknowledges the stimulating discussions with Z. Ahmed, 
S. Kailas, and S.V.S Sastry. He is also thankful to V. Jha for providing 
the reaction cross section data for some of the systems.

\end{document}